\documentclass{JHEP3}

\newcommand{\xo}{x_1}

\newcommand{\xit}{x_i}

\newcommand{\xk}{x_k}
\newcommand{\xn}{x_n}

\newcommand{\thmn}{\theta\,^{\mu\nu}}
\newcommand{\thzi}{\theta\,^{0i}}

\newcommand{\Ps}{\Psi}
\newcommand{\Psz}{\Psi_0}

\newcommand{\vf}{\varphi }
\newcommand{\vfx}{\varphi\,(x) }

\newcommand{\vff}{\varphi_f }
\newcommand{\vffo}{\varphi_{f_1}}
\newcommand{\vffd}{\varphi_{f_2}}

\newcommand{\vffp}{\varphi_{f_1}\,\star  \cdots \star \,\varphi_{f_n}}
\newcommand{\vffpts}{\varphi_{f_1}\,\ts  \cdots \ts \,\varphi_{f_n}}
\newcommand{\vffkp}{\varphi_{f_k}\,\ts  \cdots \ts \,\varphi_{f_1}}

\newcommand{\vffkpop}{\varphi_{f_{k + 1}}\,\ts  \cdots \ts \,\varphi_{f_n}}

\newcommand{\vfxtsp}{\varphi \,(x_1)\, \star\, \cdots   \star\, \varphi\,(x_n) }

\newcommand{\wfts}{W_{ \star}\,(x_1, x_2,  \ldots, x_n)}

\newcommand{\Pszl}{\langle \Psi_0}
\newcommand{\Pszr}{\Psi_0 \rangle}

\newcommand{\F}{{\Phi}}

\newcommand{\fsz}{f_0}
\newcommand{\fso}{f_1}
\newcommand{\fsd}{f_2}
\newcommand{\fsi}{f_i}
\newcommand{\fsk}{f_k}
\newcommand{\fsn}{f_n}

\newcommand{\fsx}{f\,(x)}

\newcommand{\fsoxo}{f_1\,(x_1)}
\newcommand{\fsdxd}{f_2\,(x_2)}
\newcommand{\fsixi}{f_i\,(x_i)}
\newcommand{\fsipoxipo}{f_{i + 1}\,(x_{i + 1})}

\newcommand{\fsnxn}{f_n\,(x_n)}
\newcommand{\fsdxoxd}{f_2\,(x_1, x_2)}
\newcommand{\fsnxoxn}{f_n\,(x_1, \ldots  x_n)}

\newcommand{\fsbxp}{f_1\,(x_1)\,\ts  \cdots   \ts \,{f_n\,(x_n)}}

\newcommand{\fsy}{f\,(y)}
\newcommand{\ts}{ \star }

\newcommand{\dxo}{d\,x_1}

\newcommand{\dxn}{d\,x_n}

\newcommand{\wf} {W\,(x_1,  \ldots, x_n)}

\newcommand{\reali}{\hbox{\rm I\hskip-2pt\bf R}}
\newcommand{\naturali}{\hbox{\rm I\hskip-2pt\bf N}}
\newcommand{\complessi}{\hbox{\rm I\hskip-5.9pt\bf C}}

\title{Test Functions Space\\ in Noncommutative
Quantum Field Theory}

\author{M. Chaichian$^a$,  M. Mnatsakanova$^b$, A. Tureanu$^a$
 and Yu.~ Vernov$^c$\\
$ ^a$Department of Physics, University of Helsinki and Helsinki
Institute of Physics,\\ P.O. Box
64, 00014 Helsinki, Finland\\

$^b$Skobeltsyn Institute of Nuclear Physics, Moscow State
University, Moscow, Russia \\

$^c$Institute for Nuclear Research, Russian Academy of Sciences,
Moscow, Russia }

\abstract{It is proven that the $\star$-product of field operators
implies that the space of test functions in the Wightman approach to
noncommutative quantum field theory is one of the Gel'fand-Shilov
spaces $S^{\beta}$ with $\beta < 1/2$. This class of test functions
smears the noncommutative Wightman functions, which are in this case
generalized distributions, sometimes called hyperfunctions. The
existence and determination of the class of the test function spaces
in NC QFT is important for any rigorous treatment in the Wightman
approach.}

\keywords{Noncommutative quantum field theory, axiomatic approach,
test functions spaces}

\begin{document}

\section{Introduction}
Quantum field theory (QFT) as a mathematically consistent theory was
formulated in the framework of the axiomatic approach in the works
of Wightman, Jost, Bogoliubov, Haag and others (\cite{SW} -
\cite{Haag}). Noncommutative quantum field theory (NC QFT), as one
of the generalizations of standard QFT, has been intensively
developed during the recent years (for a review, see \cite{DN}). The
idea of such a generalization of QFT ascends to Heisenberg and it
was first put forward in \cite{Snyder}. The present development in
this direction is connected with the construction of noncommutative
geometry \cite{Connes} and new physical arguments in favour of such
a generalization of QFT \cite{DFR}. Essential interest in NC QFT is
also connected with the fact that in some cases it is obtained as a
low-energy limit from the string theory \cite{SeWi}. The simplest
and at the same time most studied version of noncommutative theory
is based on the Heisenberg-like commutation relations between
coordinates,
\begin{equation}\label{cr}
[\hat{x}_\mu,\hat{x}_\nu]=i\,\theta_{\mu\nu},
\end{equation}
where $\theta_{\mu\nu}$ is a constant antisymmetric matrix.

NC QFT can be formulated also in commutative space by replacing the
usual product of operators by the star (Moyal-type) product:
\begin{equation} \label{mprodx}
\vf (x) \star\vf (x) = \exp{\left ({\frac{i}{2} \, \theta^{\mu\nu}
\, \frac{ {\partial}}{\partial x^{\mu}} \, \frac{
{\partial}}{\partial y^{\nu}}} \right)} \,\vf (x) \vf (y)|_{x=y}.
\end{equation}
This product of operators can be extended to the corresponding
product of operators in different points as well as for an arbitrary
number of operators:
$$
\vf (x_1) \star \cdots \star \vf (x_n) = \prod_{a<b} \,\exp{\left
({\frac{i}{2} \, \theta^{\mu\nu} \, \frac{ {\partial}}{\partial
x^{\mu}_a} \, \frac{ {\partial}}{\partial x^{\nu}_b}} \right)}
\,\vf (x_1)  \ldots  \vf (x_n);
$$
\begin{equation} \label{mprod}
a, b = 1,2, \ldots, n.
\end{equation}

Let us stress that actually the field operator given at a point is
not a well-defined operator (see \cite{SW} - \cite{BLOT}).
Well-defined operators are the smoothed operators:
\begin{equation} \label{vff}
\vff \equiv\int \,\vfx \,\fsx \, d \, x,
\end{equation}
where $\fsx$ is a test function. In QFT the standard assumption is
that $\fsx$ are test functions of tempered distributions.
Nevertheless in a series of papers (see \cite{Sol} - \cite{Br} and
references therein), the axiomatic approach to QFT was developed
for other spaces of test functions.

Wightman approach in NC QFT was formulated in \cite{AGM},
\cite{CMNTV} (see also \cite{VM05}). For a theory described by the
Hermitian field $\vf(x)$ and with the vacuum state denoted by
$\Psz$, the Wightman functions can be formally written down as
follows :
\begin{equation}\label{star-wightman} \label{wfg}
 \wfts = \Pszl, \vfxtsp \Pszr.
\end{equation}
The formal expression (\ref{wfg}) actually means that the scalar
product of the vectors $\F = \vffkp \, \Psi_0$ and $\Ps = \vffkpop
\, \Psi_0$ is the following:
\begin{eqnarray}\label{scprod}
\langle \,\F, \Ps \,\rangle &=& \langle \, \Psi_0, \vffpts  \,
\Psi_0\,\rangle\\
&=&\int \langle \, \Psi_0, \vf (x_1) \star \cdots \star \vf (x_n)\,
\Psi_0\,\rangle f_1(x_1)\cdots f_n(x_n)\,dx_1\cdots dx_n\cr
&=& \int \wfts\, f_1(x_1)\cdots f_n(x_n)\,dx_1\cdots dx_n \cr
&=&\int\, \wf \, \fsbxp \, \dxo \ldots \dxn\nonumber,
\end{eqnarray}
where $\wf \,
=\langle\Psi_0,\varphi(x_1)\cdots\varphi(x_n)\Psi_0\rangle$ and the
last equality is achieved by repeatedly using the definition of the
derivative of a distribution,
$$
\int \vf'(x)f(x)\, dx=-\int\vf (x)f'(x)\, dx.
$$
This choice of the product of operators $\vffo$ and $\vffd$ is
compatible with the twisted Poincar\'{e} invariance of the theory
\cite{CPT,CPrT} and also reflects the natural physical assumption,
that noncommutativity should change the product of operators not
only in coinciding points, but also  in different ones. This follows
also from another interpretation of the Heisenberg-like commutation
relations in NC QFT in terms of a quantum shift operator
\cite{CNT05}. Remark that although the expression $\wf \,
=\langle\Psi_0,\varphi(x_1)\cdots\varphi(x_n)\Psi_0\rangle$ looks
exactly like in the commutative case, the fields $\varphi(x_i)$ are
Heisenberg fields, in this case noncommutative ones, and they carry
all the characteristics of noncommutative interaction, e.g.
nonlocality in the noncommutative directions and broken Lorentz
invariance. The noncommutative Wightman functions are, however,
twisted Poincar\'e scalars (while the noncommutative Heisenberg
fields are twisted Poincar\'e covariants).

The aim of this paper is to find the class of test functions, for
which the $\star$-product (\ref{mprodx}), with the extension
(\ref{mprod}), is well defined. We should point out that, besides
the definition (\ref{mprodx}) of the Moyal $\star$-product, another
form exists, the so-called "integral representation". The two forms
are not entirely equivalent. The integral representation, or
"nonperturbative" definition of the Moyal product is the correct one
in the Moyal treatment of Quantum Mechanics. Based on it, a rigurous
study of the algebras of distributions in Quantum Mechanics was done
in \cite{GBV} and rigurous relations between the integral
representation and the "asymptotic expansion" (formally given by
(\ref{mprodx})) of the Moyal $\star$-product were established in
\cite{EGBV}. In NC QFT, it is the asymptotic expansion of the
$\star$-product which is fundamental, especially if we think in
terms of the twisted Poincar\'e symmetry, which leads naturally to
the form (\ref{mprodx}), with the extension (\ref{mprod}), of the
$\star$-product (see \cite{CPT}). Since we are interested in an
axiomatic formulation of NC QFT with twisted Poincar\'e symmetry,
our purpose in this paper is to study under which general conditions
(i.e. for which space of test functions) the asymptotic series
implied by (\ref{mprodx}) converges. It will turn out that the space
of test functions which insures the convergence of the asymptotic
expansion is more restrictive than the one found for the
nonperturbative case (integral representation). In what follows, by
$\star$-product we shall understand eq. (\ref{mprodx}) with the
extension (\ref{mprod}).

We shall prove that in order for the $\star$-multiplication to be
well defined for the functions $\fsixi$, it is necessary and
sufficient that
\begin{equation} \label{gsh}
\fsixi \in S^\beta, \qquad \beta <  1 / 2.
\end{equation}
$S^\beta$ is a Gel'fand-Shilov space \cite{GSh}. The case $\beta =
1/2$ is not excluded, but the corresponding constant $B$ (see ineq.
(\ref{gshcono})) has to be sufficiently small. We also show that
after the $\star$-multiplication we obtain functions which belong to
the same space $S^\beta$ with the same $\beta$ as $\fsixi$. In other
words, we prove that eq. (\ref{scprod}) implies that
\begin{equation} \label{new}
\langle \,\F, \Ps \,\rangle =
 \int\, \wf \, f_{\star}\,(\xo, \ldots, \xn)\, \dxo \ldots
\dxn,
\end{equation}
where $f_{\star}\,(\xo, \ldots, \xn) \equiv \fsoxo \star \fsdxd
\star \ldots \star \fsnxn \,\in S^\beta, \; \beta < 1/2$.

First we consider the case of space-space noncommutativity ($\thzi =
0$). This case is free from the problems with causality and
unitarity \cite{GM} - \cite{CNT} and in this case the main axiomatic
results: CPT and spin-statistics theorems, Haag's theorem remain
valid  \cite{AGM} - \cite{VM05}, \cite{CPrT},
\cite{CNT}-\cite{CMTV}. Then we show that all calculations can be
easily extended to the general case $\thzi \neq 0$ and moreover the
obtained results remain true in the general case as well.

These results are crucial for the derivation of the reconstruction
theorem in NC QFT. This problem will be considered in a future work.

\section{Basic Statements}

Let us point out that (as in the commutative case) the operator
$\vff$ acts on the space $J$, which is a span of all sequences of
the type:
\begin{equation} \label{vec}
g = \{ \fsz, \fsoxo, \fsdxoxd, \ldots ,\fsnxoxn \},
\end{equation}
where $\fsz \in \complessi, \, \fsk \, (\xo, \ldots \xit,  \ldots
\xk)$ is a function of $k$ variables, $\xit \in {\reali}^4$.
The sum of vectors and their multiplication by complex numbers are
defined in the standard way \cite{SW} - \cite{BLOT}.

By definition, the operator $\vff$ is determined as follows
$$
\vff \, \{ \fsz, \fso, \fsd \ldots \fsn \} = \{ f \fsz, f \star
\fso, f \star \fsd, \ldots ,f \star \fsn \},
$$
\begin{equation} \label{vff}
f \equiv \fsx, \quad \fsk \equiv \fsk\,(\xo, \ldots \xk), \quad f
\star \fsk \equiv f \, \exp{\left ({\frac{i}{2} \, \theta^{\mu\nu} \,
\frac{\overleftarrow{\partial}}{\partial x^{\mu}} \, \frac{\overrightarrow{
\partial}}{\partial x^{\nu}_1}} \right)} \, \fsk.
\end{equation}
The special set of vectors
\begin{equation} \label{vc}
\{\Psi_0, \, \vffo  \Psi_0, \, \ldots \, ,\vffp  \, \Psi_0\},
\end{equation}
where $\Psi_0 \equiv \{1, 0, \ldots 0 \}, \; \fsi \equiv \fsixi, \,
\xit \in \reali^4$, forms a dense domain $J_0$ in the space under
consideration. The scalar product in $J_0$ is defined by Wightman
functions (see eq. (\ref{scprod})).

The necessary condition
\begin{equation} \label{14}
\langle \,\F, \Ps \,\rangle = \overline{\langle \, \Psi, \F
\,\rangle}
\end{equation}
is satisfied (just as in the commutative case) if
\begin{equation} \label{wfe}
W_\star(x_1,\cdots,x_n) = \overline{W_\star(x_n,\cdots,x_1)}.
\end{equation}
In fact, due to the antisymmetry of $\thmn$, condition
(\ref{wfe}) leads to (\ref{14}).

Let us recall that the scalar product between any two vectors in $J$
is defined by a finite linear combination of Wightman functions with an
arbitrary accuracy. This fact is crucial in the derivation of
axiomatic results both in commutative and noncommutative cases.

Our aim is to determine the spaces on which eq. (\ref{scprod}), that
is the $\star$-multiplication, is well-defined. Evidently the space
of tempered distributions cannot be compatible with the
$\star$-multiplication, as each function of this space admits only a
finite number of derivatives \cite{SW}.  Gel'fand and Shilov proved
that if $\fsx \in S^{\beta}$ (see ineq. (\ref{gshcono})) then the
series of derivatives of infinite order is well-defined on such a
space. Thus we assume that $\fsx \in S^{\beta}$ and prove that the
$\star$-product  is well-defined only if each $\fsi$ belongs to the
Gel'fand-Shilov space $S^\beta, \, \beta < 1/2$. The $\star$-product
is well-defined also if $\beta = 1/2$, but only for functions which
satisfy inequality (\ref{gshcono}) with sufficiently small $B$.

Let us recall the definition and basic properties of
Gel'fand-Shilov spaces $S^\beta$ \cite{GSh}. In the case of one
variable $\fsx, \, x \in \reali^1$ belongs to the space $S^\beta$,
if the following condition is satisfied:
\begin{equation} \label{gshcono}
\left |\,  x^{k} \, \frac{ {\partial}\,^{q}\,\fsx}{\partial x
^{q}} \right | \leq C_{k}\, B\,^{q}\, q\,^{q\beta}, \quad - \infty
< x < \infty, \quad k,q \in \naturali,
\end{equation}
where the constants $C_{k}$ and $B$ depend on the function $\fsx$.
Below we use the inequality (\ref{gshcono}) only at $k = 0$:
\begin{equation} \label{gshcond}
\left |\, \frac{ {\partial}\,^{q}\,\fsx}{\partial x ^{q}} \right |
\leq C \, B\,^{q}\, q\,^{q\beta}, \quad - \infty < x < \infty,
\quad q \in \naturali.
\end{equation}
In the case of a function of several variables, the inequality
(\ref{gshcond}) holds for any partial derivative:
\begin{equation} \label{gshcons}
\left |\, \frac{ {\partial}\,^{q}\,f\,(x^{1}, \ldots
x^{k})}{{(\partial x ^{i})}^{q}} \right | \leq C \, B\,^{q}\,
q\,^{q\beta}, \quad - \infty < x_{i} < \infty, \quad q \in
\naturali.
\end{equation}
As our results do not depend on the constant $C$, in what follows we
put $C = 1$.

\section{Proof of the Statement}

We point out that if the $\star$-product is well-defined for
$\fsixi\,\star\,\fsipoxipo$, it is also well-defined for all
expressions in the right-hand side of eq. (\ref{scprod}).

Let us stress that, in fact, under the same conditions
$$
\exp{\left ({\frac{i}{2} \, \theta^{\mu\nu} \, \frac{
{\partial}}{\partial x^{\mu}} \, \frac{ {\partial}}{\partial
y^{\nu}}} \right)} \,f\,(x, y)
$$
is well-defined as well, if $f\,(x, y) \in S^{\beta}, \, \beta <
1/2$.

First we consider the case when the $\star$-multiplication acts only
on the coordinates $x_i^1$ and  $x_i^2$, i.e. the case of
space-space noncommutativity with only $\theta^{12}=-\theta^{21}\neq
0$. Then we show that the proof can be extended to the general case
$\thzi \neq 0$ and the corresponding results are the same.

Let us study
\begin{equation} \label{fxsfy}
f\,(x)\,\star\,f\,(y) = \exp{\left ({\frac{i}{2} \,
\theta^{\mu\nu} \, \frac{ {\partial}}{\partial x^{\mu}} \,
\frac{{\partial}}{\partial y^{\nu}}} \right)} \,f\,(x)\,f\,(y).
\end{equation}
We have to find the conditions under which the series
\begin{equation} \label{serfxsfy}
\sum^{\infty}_{n = 0}\,\frac{1}{n!}\,{\left ({\frac{i}{2} \,
\theta^{\mu\nu} \, \frac{ {\partial}}{\partial x^{\mu}} \,
\frac{{\partial}}{\partial y^{\nu}}} \right)}^{n} \,f\,(x)\,f\,(y)
\equiv \sum^{\infty}_{n = 0}\,\frac{D_{n}}{n!}
\end{equation}
converges. Evidently
$$
D_{n} = \sum_{n_{\mu}, n_{\nu}}
{\left (\frac{i}{2} \theta\,^{\mu\nu} \right)}^{n_{\mu}}
{\left (\frac{i}{2} \theta\,^{\nu\mu} \right)}^{n_{\nu}}
\frac{{\partial}\,^{n_{\mu}}}{({\partial x^{\mu})}^{n_{\mu}}}
\frac{{\partial}\,^{n_{\mu}}}{({\partial y^{\nu})}^{n_{\mu}}}
\frac{{\partial}\,^{n_{\nu}}}{({\partial x^{\nu})}^{n_{\nu}}}
\frac{{\partial}\,^{n_{\nu}}}{({\partial y^{\mu})}^{n_{\nu}}}
\,f\,(x)\,f\,(y),
$$
\begin{equation} \label{serdn}
\mu, \nu = 1, 2, \quad n_{\mu} + n_{\nu} = n.
\end{equation}
We show that actually it is sufficient to estimate one term in the
series (\ref{serdn}), that is to estimate the module of
\begin{equation} \label{estim}
{\theta}\,^{n}\,\frac{{\partial}\,^{n_{\mu}}}{({\partial
x^{\mu})}^{n_{\mu}}} \, \frac{{\partial}\,^{n_{\mu}}}{({\partial
y^{\nu})}^{n_{\mu}}}\, \frac{{\partial}\,^{n_{\nu}}}{({\partial
x^{\nu})}^{n_{\nu}}} \, \frac{{\partial}\,^{n_{\nu}}}{({\partial
y^{\mu})}^{n_{\nu}}}\, \,f\,(x)\,f\,(y) \equiv
{\theta}\,^{n}\,B\,(n_{\mu},n_{\nu}), \ \ \theta \equiv \left|
\frac{i\,\theta\,^{\mu\nu}}{2}\right|.
\end{equation}
Using the inequality (\ref{gshcons}) and taking into account that
$n_{\mu}, n_{\nu} \leq n$, we obtain
\begin{equation} \label{bn}
|B\,(n_{\mu},n_{\nu})| < B\,^{2\,(n_{\mu} +
n_{\nu})}\,{n_{\mu}}^{2\beta{n_{\mu}}}\,{n_{\nu}}^{2\beta{n_{\nu}}}
< n^{2\beta\,({n_{\mu}} + n_{\nu})}B\,^{2n} =
n^{2\beta\,n}\,B\,^{2n},
\end{equation}
as $n_{\mu} + n_{\nu} = n$. This estimate  is one and the same for
any $n_{\mu}, n_{\nu}$. As the total number of these terms in
$D_{n}$ is $2^{n}$, taking into account that  the module of the sum
is less than the sum of the moduli, we come to the following
inequality
\begin{equation} \label{dn}
|D_{n}| < {(2\,\theta\,B^{2})}^{n}\,n^{2\,n\,\beta}.
\end{equation}
Using this  inequality and the fact that, according to the Stirling
formula, $\frac{1}{n!} < {\left( \frac{e}{n} \right)}^{n}$, we come
to the estimate
\begin{equation} \label{dnsec}
\left|\frac{D_{n}}{n!}\right| < {\tilde{B}}^{n}\,n^{-
2\,n\,\gamma},
\end{equation}
where $\tilde{B} = 2\,e\,\theta\,B^{2}, \; \gamma = 1 /2-
\beta$.\\
For any $\tilde{B}$ the series
\begin{equation} \label{m}
\sum^{\infty}_{n = 0}{\tilde{B}}^{n}\,n^{- 2\,n\,\gamma}
\end{equation}
converges if  $\gamma > 0$, i.e. $\beta < 1/2$, and diverges if
$\beta
> 1/2$. If  $\beta = 1/2$ the series converges if $\tilde{B} < 1$.

Thus we come to the conclusion that the series (\ref{serfxsfy}) for
arbitrary $B$ and $C$ is convergent if $\beta < 1/2$ and divergent
if $\beta > 1/2$. If $\beta = 1/2$ the series converges at
sufficiently small $B$.

Now let us show that the function $f_{\star}\,(x, y) \equiv \fsx
\star \fsy$ belongs to the same Gel'fand-Shilov space $S^{\beta}, \,
\beta < 1/2$ as $\fsx$.

According to the inequality (\ref{gshcons}) we have to prove that
\begin{equation} \label{25}
\left| \frac{{\partial}\,^{q}}{({\partial x^{\mu})}^q}
\,f_{\star}\,(x, y)\right| < C'\,{{B'}}^{q}\,q^{q\,\beta}, \quad -
\infty < x^{\mu} < \infty
\end{equation}
with some constants $C'$ and $B'$.

As above, first we consider the corresponding estimate for one term
$B\,(n_{\mu}, n_{\nu})$ in $D_{n}$ (see eqs. (\ref{serdn}),
(\ref{estim})). Actually we have to estimate
$$
\frac{{\partial}\,^{n_{\mu} + q}}{({\partial x^{\mu})}^{n_{\mu} +
q}}\,\fsx.
$$
In accordance with eq. (\ref{gshcons})
\begin{equation} \label{26}
\left|\frac{{\partial}\,^{n_{\mu} + q}}{({\partial x^{\mu})}^{n_{\mu} +
q}}\,\fsx \right| < B^{n_{\mu} + q}\,{(n_{\mu} + q)}^{(n_{\mu} + q)\beta}.
\end{equation}
Evidently,
$$
{(n_{\mu} + q)}^{(n_{\mu} + q)\beta} =
q^{q\beta}\,{n_{\mu}}^{n_{\mu}\beta}\,{\left(1 + \frac{1}{x}
\right)}^{xq\beta}\,{\left(1 +
x\right)}^{\frac{1}{x}n_{\mu}\beta}, \quad x = \frac{n_{\mu}}{q}.
$$
After elementary calculations we obtain
\begin{equation} \label{27}
{(n_{\mu} + q)}^{n_{\mu} + q} <
q^{q\beta}\,{n_{\mu}}^{n_{\mu}\beta}\,e^{q\beta}\,e^{n_{\mu}\beta}
<e^{q\beta}\,q^{q\beta}\,e^{n\beta}\,n^{n_{\mu}\beta}.
\end{equation}
Thus according to the ineq. (\ref{26}) we have
\begin{equation} \label{28}
\left|\frac{{\partial}\,^{n_{\mu} + q}}{({\partial x^{\mu})}^{n_{\mu} +
q}}\,\fsx \right| <
{(Be^{\beta})}^{q}\,q^{q\beta}B^{n}e^{n\beta}n^{n_{\mu}\beta}.
\end{equation}
Comparing the bounds (\ref{dnsec}) and (\ref{28}), we see that
\begin{equation} \label{29}
\left|\frac{{\partial}\,^{q}}{({\partial x^{\mu})}^{q}}\,B\,(n_{\mu},
n_{\nu}) \right| <
{(Be^{\beta})}^{q}\,q^{q\beta}B^{2n}e^{n\beta}n^{n\beta}.
\end{equation}
Thus
\begin{equation} \label{30}
\frac{{\partial}\,^{q}}{({\partial
x^{\mu})}^{q}}\,\,f_{\star}\,(x, y) = \sum^{\infty}_{n =
0}\,\frac{1}{n!}\frac{{\partial}\,^{q}D_{n}}{({\partial
x^{\mu})}^{q}}
\end{equation}
admits the following estimate
\begin{equation} \label{31}
\left|\frac{{\partial}\,^{q}}{({\partial
x^{\mu})}^{q}}\,\,f_{\star}\,(x, y) \right|  <
{(Be^{\beta})}^{q}\,q^{q\beta}\,\sum^{\infty}_{n =
0}\,\frac{1}{n!}\,{(B^{2}e^{\beta})}^{n}\,n^{n\beta}.
\end{equation}
The series in the right-hand side of eq. (\ref{31}) converges if
$\beta < 1/2$ and diverges if $\beta > 1/2$ on the same grounds as
the series (\ref{m}). Thus we see that the necessary condition on
$f_{\star}\,(x, y)$ (\ref{26}) is fulfilled with
$$
B' = B\,e^{\beta} \quad \mbox{and} \quad C' = \sum^{\infty}_{n =
0}\,\frac{1}{n!}\,{(B^{2}e^{\beta})}^{n}\,n^{n\beta}.
$$

Let us proceed to the general case $\thzi \neq 0$. We prove that in
this case the $\star$-multiplication is well defined (just as in the
case of space-space noncommutativity) if the corresponding test
functions belong to $S^{\beta}, \, \beta < 1/2$.

In order to show this, let us rewrite the condition (\ref{gshcons}) in
a more general form, using the standard notations \cite{SW},
\cite{GSh}. By definition, $\fsx \in S^{\beta}, x \in
\reali^{k}$ if
\begin{equation} \label{33}
\left| \frac{D^{q}}{{\partial x}^{q}} \right| < C\,
{B}^{q}\,q^{q\beta},
\end{equation}
where
$$
x^{q} = {(x^{1})}^{q_{1}}\ldots {(x^{k})}^{q_{k}} \quad D^{q} =
\frac{ {\partial}\,^{|q|}}{{\left(\partial x^{1}\right)}^{{q}_1}
\ldots {\left(\partial x^{k}\right)}^{{q}_{k}}}, \quad |q|
=q_{1}+q_{2}+\ldots+q_k.
$$
As before we consider one term in the expansion (\ref{serdn}) (now
$\mu$ and $\nu$ possess the values $0,1,2,3$.

Since condition (\ref{33}) contains only $q$, we can use our
previous consideration for an arbitrary term in $D_{n}$ and come to
a similar estimate (but now $\theta = \max|\thmn/2|$). This estimate
leads to the corresponding one on $D_{n}$, the only difference being
that it is necessary to substitute $2^{n}$ by $4^{n}$ since $\thmn$
by rotation can always be reduced to a block-diagonal form with four
nonzero components (e.g. $\theta_{01}=-\theta_{10}=\theta$ and
$\theta_{23}=-\theta_{32}=\theta'$, all the other components being
zero). As the set (\ref{m}) converges at arbitrary $\tilde B$, if
$\beta < 1/2$, and diverges if $\beta
> 1/2$, the results obtained for space-space noncommutativity
remain true also in the general case. If $\beta = 1/2$ the words
"sufficiently small $B$" have different meanings in the cases $\thzi
= 0$ and $\thzi \neq 0$.

\section{Conclusions}

In this paper we have proven that the space of test functions for
which the Moyal $\star$-product is well-defined, in other words, the
space of test functions for the Wightman distribution functions
corresponding to the NC QFT, is one of the Gel'fand-Shilov spaces
$S^{\beta}$ with $\beta < 1/2$. This class of test functions smears
the NC Wightman functions which are generalized distribution
functions, called sometimes hyperfunctions.

The existence and determination of the class of test functions
spaces is important for any rigorous treatment of the axiomatic
approach to NC QFT via NC Wightman functions and the derivation of
rigorous results such as CPT and spin-statistics theorems, and is
also needed for the derivation of other results in axiomatic
approach such as the cluster-decomposition property of  NC Wightman
functions and eventually the proof of the reconstruction theorem.

Recall that the class of test functions in the ordinary QFT contains
functions with compact support. In the case of  NC Wightman
functions, however, the set of test functions consists of functions
with non-compact support only in the NC coordinates.

Note, however, that in the case of space-space noncommutativity,
i.e. $\thzi = 0$, the test functions can still have finite support
in the commutative directions $x_{0}, x_{3}$. As a result, the local
commutativity condition can be formulated in these directions as
\begin{equation} \label{34}
\vffo \star \vffd = \vffd \star \vffo,
\end{equation}
where the test functions $f_{1}\,(x)$ and $f_{2}\,(x')$ are zero
everywhere except on space-like separated finite domains $O$ and
$O'$ in the commutative coordinates, i.e. for each pair of points $x
\in O$ and $x' \in O'$
\begin{equation} \label{35}
{\left(x_{0} - {x_{0}}' \right)}^{2} -  {\left(x_{3} - {x_{3}}'
\right)}^{2} < 0,
\end{equation}
but without any restriction in the noncommutative directions $x_{1}$
and $x_{2}$. This is in effect the well-known light-wedge locality
condition \cite{AGM}, \cite{CMNTV}.

It should be noted that, although according to the twisted
Poincar\'e symmetry of NC QFT, the Wightman functions should be
defined with $\star$-product as in (\ref{star-wightman}), for
practical purposes one may as well define them with usual product,
because the nonlocality is taken into account by the very definition
of the Heisenberg fields \cite{AGM}. Therefore, although in the
smeared noncommutative Wightman functions the test functions will
not be multiplied with $\star$-product, they still have to belong to
the Gel'fand-Shilov space found in this paper, to account for the
character of generalized distributions of the noncommutative
Wightman functions\footnote{In a recent paper \cite{Fiore} it was
argued that due to the translational invariance of NC QFT, the
commutative and noncommutative Wightman functions are practically
the same, with similar properties, leading to the fact that
commutative and noncommutative QFT are actually identical. We argue
here that this cannot be the case, since the spaces of test
functions in the two situations are completely different,
emphasizing the deep qualitative difference between the commutative
Wightman functions (tempered distributions) and the noncommutative
Wightman functions (generalized distributions, i.e.
hyperfunctions).}.

In the general case $\thzi \neq 0$, we have also shown that the
space of test functions is the same Gel'fand-Shilov space
$S^{\beta}$ with $\beta < 1/2$, but with respect to all
noncommutative coordinates. Thus in the general case we have not the
standard condition of local commutativity. As CPT and
spin-statistics theorems have been proven for very general spaces
\cite{Sol} we can conclude that the finding of the class of the
spaces, in which $\star$-multiplication and thus noncommutative
Wightman functions are well-defined, gives rise to the possibility
to prove CPT-theorem and, maybe, also spin-statistics theorem for
the general case. It should be mentioned that the twisted Poincar\'e
symmetry provides an answer at least for the spin-statistics
relation in the case of general noncommutativity: the
spin-statistics relation holds for NC QFT with time-space
noncommutativity, provided that such theories can be consistently
defined \cite{twist review}. Previous perturbative studies have
shown however that such theories are pathological
\cite{GM}-\cite{CNT}. The axiomatic study of time-space
noncommutative theories based, among other aspects, on the space of
test functions found in this work, may resolve the problem,
indicating whether or not the time-space noncommutative theories are
well defined and the pathologies are mere artifacts of the
perturbation theory.

\section*{Acknowledgements}
We are grateful to J. M. Gracia-Bond\'ia and V. S. Vladimirov for
several enlightening discussions. Yu. S. Vernov is partly supported
by the grant of the President of the Russian Federation
NS-7293.2006.2 (Government Contract 02.445.11.7370). The support of
the Academy of Finland under the grants no. 122577 and 122596 is
acknowledged.

\end{document}